\documentclass[aps,prb,reprint,superscriptaddress,showpacs,amsmath,amssymb]{revtex4-2}
\usepackage[T1]{fontenc}
\usepackage{graphicx}
\usepackage[version=4]{mhchem}
\usepackage{textcomp}
\usepackage{dcolumn}
\usepackage{bm}
\usepackage{esdiff}
\usepackage[colorlinks=true,citecolor=blue,linkcolor=blue,urlcolor=blue,pdfhighlight =/O]{hyperref}
\usepackage{float} 
\usepackage{upgreek}
\begin{document}
	\title{Determining the absolute value of magnetic penetration depth in small-sized superconducting films}
	
	\author{Ruozhou Zhang}
	\altaffiliation{These authors contributed equally to this work.}

	\affiliation{Beijing National Laboratory for Condensed Matter Physics, Institute of Physics, Chinese Academy of Sciences, Beijing 100190, China}
	\affiliation{School of Physical Sciences, University of Chinese Academy of Sciences, Beijing 100049, China}
	
	\author{Mingyang Qin}
	\altaffiliation{These authors contributed equally to this work.}
	\affiliation{Beijing National Laboratory for Condensed Matter Physics, Institute of Physics, Chinese Academy of Sciences, Beijing 100190, China}
	\affiliation{School of Physical Sciences, University of Chinese Academy of Sciences, Beijing 100049, China}
	
	\author{Lu Zhang}
	\affiliation{Shanghai Institute of Microsystem and Information Technology, Chinese Academy of Sciences, Shanghai 200050, China}
	
	\author{Lixing You}
	\affiliation{Shanghai Institute of Microsystem and Information Technology, Chinese Academy of Sciences, Shanghai 200050, China}
	
	\author{Chao Dong}
	\affiliation{Institute of High Energy Physics, Chinese Academy of Sciences, Beijing 100049, China}
	
	\author{Peng Sha}
	\affiliation{Institute of High Energy Physics, Chinese Academy of Sciences, Beijing 100049, China}
	
	\author{Qihong Chen}
	\affiliation{Beijing National Laboratory for Condensed Matter Physics, Institute of Physics, Chinese Academy of Sciences, Beijing 100190, China}
	
	\author{Jie Yuan}
	\affiliation{Beijing National Laboratory for Condensed Matter Physics, Institute of Physics, Chinese Academy of Sciences, Beijing 100190, China}
	\affiliation{Songshan Lake Materials Laboratory, Dongguan, Guangdong 523808, China}
	
	\author{Kui Jin}
	\altaffiliation{Corresponding author: \textcolor{blue}{kuijin@iphy.ac.cn}}
	\affiliation{Beijing National Laboratory for Condensed Matter Physics, Institute of Physics, Chinese Academy of Sciences, Beijing 100190, China}
	\affiliation{School of Physical Sciences, University of Chinese Academy of Sciences, Beijing 100049, China}
	\affiliation{Songshan Lake Materials Laboratory, Dongguan, Guangdong 523808, China}
	
	
	\date{\today}
	
	\begin{abstract}
		In the previous four decades, a two-coil mutual inductance (MI) technique has been widely employed in characterizing magnetic penetration depth, $\lambda$, of superconducting films. However, the conventional methods used to obtain $\lambda$ are not applicable to small-sized films with common shapes, which limits the application of the MI technique in superconductivity research. Here, we first employed the fast wavelet collocation (FWC) method to a two-coil system and then proposed the possibility of directly obtaining the absolute $\lambda$ of polygonal superconducting films with arbitrary sizes. To verify its accuracy and advantages, we extracted the $\lambda$ values  of square NbN films with different sizes using the FWC and conventional flux leakage subtraction (FLS) methods.  Notably, the FLS method fails for a $5\times 5 \, \rm mm^2$ film, which is attributed to the significant current peak at the film edge. In contrast, the absolute $\lambda$ extracted using the FWC method was independent of the film size. Finally, we established the applicability of the FWC method to large coil spacings, which may pave the way for integrating high-accuracy $\lambda$ measurements with the ionic liquid gating technique.
	\end{abstract}
	
	
	\maketitle
	\section{Introduction}
	As a key physical parameter for superconductors, the magnetic penetration depth, $\lambda$, links macroscopic electrodynamics with the microscopic mechanism of superconductivity \cite{Tinkham}. First, $ \lambda^{-2} $ is proportional to the superfluid density, $n_s$, and its temperature dependence, $ \lambda^{-2} (T)$, encodes the information on paring symmetry and multiband superconductivity \cite{Hardy-1993,Skinta-2002,Fletcher-2005}. Second,  by further extrapolating to the zero-temperature limit, the superfluid phase stiffness, $\rho_{\rm s0}\propto\lambda^{-2} (T\to 0)$, could be extracted, which represents the resiliency of the superconducting phase under quantum or thermal fluctuations \cite{Uemura-1989,Emery-1995,Hetel-2007,Broun-2007,Franz-2006}. Third, according to the London phenomenological model, $ \lambda^{2} $ is proportional to the effective mass, $m^*$, which could directly reflect the impact of a quantum critical point \cite{Hashimoto-2012,WangCG-2018,Joshi-2020}. In addition, the performances of most applicable superconducting devices depend on $\lambda$, e.g., the surface resistance, $R_{\rm s}$, of microwave filters \cite{Gurevich-2017} and the superheating field, $B_{\rm sh}$, of radio-frequency cavities in accelerators \cite{Kubo-2017,Lin-2021}. That is, high-precision measurement of the absolute value of  $\lambda$  is crucial for elucidating the mechanism of superconductivity and exploring the applications of superconductors.
	\par 
	However, measuring the absolute value of $\lambda$ accurately is difficult because $\lambda$ is on the order of thousands of angstroms. Currently, scientists have developed various techniques \cite{Prozorov-2006,Prozorov-2011}, among which the two-coil mutual inductance (MI) technique is of particular interest. Owing to its simplicity, lack of destruction, and high sensitivity, the technique has been used to characterize $\lambda$ in a wide range of superconducting films. Over the previous 40 years, the MI technique has provided insight into the nature of superconductivity, including the Berezinskii-Kosterlitz-Thouless transition in Al and NbN thin films \cite{Hebard-1980,Kamlapure-2010}, quantum criticality in strongly underdoped $\rm Y_{1-x}Ca_xBa_2Cu_3O_{7- \delta}$ ultrathin films \cite{Hetel-2007}, and the scaling law between $\rho_{\rm s0}$ and critical transition temperature, $T_{\rm c}$, in $\rm La_{2-x}Sr_xCuO_4$ films \cite{Bozovic-2016}. Recently, Jia \emph{et al.} \cite{Jia-2019} reported a dome-shaped superconducting region in K-absorbed FeSe films using an \emph{in situ} MI device in a multifunctional scanning tunneling microscope, demonstrating the advantage of the MI technique in characterizing fragile samples.
	\par
	In general, a MI device consists of a drive coil and a pickup coil, which are coaxially located on the same side (\emph{reflection}-type) or opposite sides (\emph{transmission}-type) of the superconducting film. When the film enters the Meissner state, the magnetic field produced by the alternating current in the drive coil is expelled by the induced screening current in the film. Consequently, the pickup coil voltage, $V$, or equivalent mutual inductance, $M$, undergoes an instantaneous change, from which the absolute value of $\lambda$ can be extracted \cite{Claassen-1997,Jeanneret-1989,Turneaure-1996,Turneaure-1998}.
	\par
	For the \emph{reflection}-type MI setup, it is easy to implement further manipulations on the film \cite{Jia-2017,Nam-2018,Kinney-2015} because both coils are under the substrate and the top side of the film is free. However, it is difficult to achieve high-precision measurement of  $\lambda$ when using this configuration. This is because there is no method for eliminating the errors arising from the uncertainties in the coil geometry, most of which are owing to the nonideal aspects of coil windings and thermal shrinkage when the sample is cooled down. In contrast, for the \emph{transmission}-type MI setup, these uncertainties can be removed by normalizing the measured mutual inductance, $M_{\rm exp}$, to its normal state value, $M_{\rm exp} (T>T_{\rm c})$ \cite{Turneaure-1996,Turneaure-1998}. Moreover, when the radius of the film is infinite, the normalized mutual inductance, $M_{\rm exp}/M_{\rm exp} (T>T_{\rm c})$, can be expressed analytically; the expression was first derived by Clem \emph{et al.} \cite{Clem-1992}. In practice, the finite size of the film allows some magnetic flux “leaks” around the film edge, thus resulting in residual coupling, $M_1$. Notably, it was established that $M_1$ is independent of $\lambda$ and depends only on the shape of the film \cite{Turneaure-1998,He-2016}; thus it can be evaluated experimentally by substituting a thick Nb film with the same shape as the sample. Thereafter, $\lambda$ can be extracted from the “corrected” mutual inductance $(M_{\rm exp}-M_1)/M_{\rm exp} (T>T_{\rm c})$. This method has been widely used, and it is referred to as the flux leakage subtraction (FLS) method in this study.
	\par
	However, the FLS method preserves high precision only for small $M_1/M_{\rm exp} (T>T_{\rm c})$ \cite{Turneaure-1996,Turneaure-1998}, which intuitively requires a large film size. Therefore, Fuchs \emph{et al.} \cite{Fuchs-1996} suggested a film with a diameter greater than 50 mm for accuracy. However, the preparation of high-quality large-sized superconducting films is challenging. Although the conventional numerical model is independent of the film size, it can only deal with circular films \cite{Turneaure-1996}. For small-sized films with common shapes such as squares, an accurate method for extracting the absolute $\lambda$ from $M$ is required.
	\par
	In this study, we first employed the fast wavelet collocation (FWC) method to extract the absolute $\lambda$ from the mutual inductance data, which in principle applies to polygonal superconducting films with arbitrary sizes. Additionally, the details of the numerical model are presented. To evaluate the accuracy and advantages, we compared the values of $\lambda$ obtained using the FWC method with those obtained using the FLS method. 
	\section{Experimental Methods}
	The inset of Fig.~\ref{fig1} shows a schematic of our \emph{transmission}-type MI device, in which the drive and pickup coils are coaxially located on opposite sides of the film. Both coils were wound using $40 \,   \upmu \rm m$ oxygen-free copper (OFC) wires with insulation coating. Their inner diameter was 0.5 mm, the outer diameter was 1.3 mm, and the length was 1.6 mm. The separation between the two coils was approximately 1 mm. The device was thermally connected to a 3 K platform of a Montana Instruments cryocooler. The drive current had a frequency of 10 kHz and an amplitude of 2 mA, supplied by a Stanford Research SR830 lock-in amplifier. The induced voltage, $V=V_x+iV_y$, in the pickup coil was measured using the same lock-in amplifier with a reference phase of $90^\circ$. More details can be found in our previous work \cite{Zhang-2020}.
	\par
	The mutual inductance M of the two coils can be determined as
	\begin{eqnarray}
		M=\frac{V_x}{\omega I_d}+i\frac{V_y}{\omega I_d}
		\label{eq:1},
	\end{eqnarray}
	where $\omega$ is the angular frequency and $I_d$ is the amplitude of the drive current. The first term in Eq.~(\ref{eq:1}) represents inductive coupling, whereas the second represents resistive coupling. Except at temperatures near $T_{\rm c}$, the film response is purely inductive and the resistive coupling is negligible \cite{Turneaure-1996}. 
	\par 
	We fabricated high-quality NbN films via reactive DC magnetron sputtering, as detailed in \cite{Zhang2-2018}. Films with sizes of  $10\times 10 \, \rm mm^2$(NbN\#1) and  $5\times 5 \, \rm mm^2$ (NbN\#2) were grown in the same branch. The substrates were $(100)$-oriented MgO single crystals. The thickness of the films was $6.5\pm 0.2\rm \, nm$, as characterized by X-ray reflectivity.
	\section{\label{sec:level3}Numerical Model}
	The role of our numerical model is to establish a one-to-one correspondence between $\lambda$ and $M$. In principle, two types of currents contribute to mutual inductance $M$. One is the alternating current in the drive coil. The other is the screening current in the superconducting film, which needs to be carefully determined.
	\par 
	We consider a polygonal superconducting film placed on the  $xy$ plane. The thickness of the film is $d$, and the projection of the film on the $xy$ plane is a polygon $\Omega$. We assume that the vector potential, ${\bf A_d}=A_{dx}\hat{x}+A_{dy}\hat{y}$, generated by the drive current and screening current density ${\bf j_s}=j_{sx}\hat{x}+j_{sy}\hat{y}$ in the film are parallel to the film surface. Thereafter, employing London’s and Maxwell’s equations, $j_{s\alpha}\, (\alpha=x,y)$ is given by \cite{Turneaure-1996,Turneaure-1998}
	
	\begin{eqnarray}
		j_{s\alpha}({\bf r})+\frac{d_{\rm eff}}{4\pi \lambda^2}\int_{\Omega}^{}d^2{\bf r}'\frac{j_{s\alpha}({\bf r}')}{|{\bf r}-{\bf r}'|}=-\frac{1}{\mu_0 \lambda^2}A_{d\alpha}({\bf r})
		\label{eq:2},
	\end{eqnarray}
	where ${\bf r}=x{\bf \hat{x}}+y{\bf \hat{y}}$ is the coordinate in $\Omega$, $d_{\rm eff}=\lambda \sinh(d/\lambda)$ is the effective thickness, and $\mu_0=4\pi\times 10^{-7} \rm N A^{-2}$ is the vacuum permeability. 
	\par
	Eq.~(\ref{eq:2}) is a two-dimensional Fredholm integral equation of the second kind with a weakly singular kernel. This type of equations are of great importance in various engineering application fields \cite{Chen-2002}, while their solution is an unresolved problem until the 21st century. We propose to solve Eq.~(\ref{eq:2}) by the fast wavelet collocation (FWC) algorithm, which was first developed by Chen \emph{et al.} in 2002 \cite{Chen-2002}. Its significant computational efficiency and attractive convergence properties have been demonstrated in   \cite{Chen-2002,Xu-2005,Chen-2008}. In the rest of this section, we will take the rectangular film as an example to describe the calculation steps of the FWC method.
	\par 
	First, we subdivide the polygon $\Omega$ into several triangles, and there is at most one common edge or vertex for two different triangles. A rectangular film with a length of $2a$ and a width of $2b$ can be divided into two triangles:
	$\Delta_0=\{(x,y)\in \mathbb R^2:x\geq -a,y\leq b,ay-bx\geq 0 \}$ and $\Delta_1=\{(x,y)\in \mathbb R^2:x\leq a,y\geq -b,ay-bx< 0 \}$. Because $\Delta_0$ and $\Delta_1$ can be affinely mapped onto the unit triangle $E=\{(x,y)\in \mathbb R^2:0\leq x\leq y\leq 1\}$, Eq.~(\ref{eq:2}) can be rewritten as
	\begin{align}
		\widetilde{j_{s\alpha}}^1(x,y)+\frac{d_{\rm eff}}{2\pi \lambda^2}&
		\left [
		b(K\widetilde{j_{s\alpha}}^1)\left(x,y,\frac{b}{a}\right) \right.\nonumber\\
		+&
		\left. a
		(K\widetilde{j_{s\alpha}}^2)\left(y,x,\frac{a}{b}\right)
		\right ] 
		=
		f_\alpha(x,y),\nonumber
	\end{align}
	\begin{align}
		\widetilde{j_{s\alpha}}^2(x,y)+\frac{d_{\rm eff}}{2\pi \lambda^2}&
		\left [
		b(K\widetilde{j_{s\alpha}}^1)\left(y,x,\frac{b}{a}\right) \right.\nonumber\\
		+&
		\left. a
		(K\widetilde{j_{s\alpha}}^2)\left(x,y,\frac{a}{b}\right)
		\right ] 
		=
		f_\alpha(y,x)
		\label{eq:3},
	\end{align}
	where $\widetilde{j_{s\alpha}}^1(x,y)=j_{s\alpha}(2ax-a,2by-b)$, $\widetilde{j_{s\alpha}}^2(x,y)=j_{s\alpha}(2ay-a,2bx-b)$, $f_{\alpha}(x,y)=-\frac{1}{\mu_0\lambda^2}A_{d\alpha}(2ax-a,2by-b)$, and the effect of operator $K$ is $(KF)(x,y,t)=\int_{E}^{} dx'dy'\frac{F(x',y')}{\sqrt{(x'-x)^2+t^2(y'-y)^2}}$.
	\par 
	Next, we expand $\widetilde{j_{s\alpha}}^\beta \, (\alpha=x,y, \, \beta=1,2)$ using multi-scale wavelets $\{\omega_{ij} \}$(see Appendix \ref{sec:appendix A1}) as 
	\begin{equation}
		\widetilde{j_{s\alpha}}^\beta(x,y)=\sum_{i=0}^{n}\sum_{j=0}^{\omega(i)}u_{ij\alpha}^\beta \omega_{ij}(x,y)
		\label{eq:4},
	\end{equation} 
	where $i$ and $j$ are integers, $\omega(0)=3$, $\omega(i)=9\times 4^{i-1}\, (i\geq 1)$, and $n\geq 1$ is the highest level of $\omega_{ij}$ (for rectangular films, we take $n=6$ \cite{Xu-2005}). Thereafter, by substituting Eq.~(\ref{eq:4}) into Eq.~(\ref{eq:3}) and applying collocation functional $\ell_{i'j'}$ (see Appendix \ref{sec:appendix A2}) on both sides of Eq.~(\ref{eq:3}), we obtain the matrix equation
	\begin{align}
		\sum_{i=0}^{n}\sum_{j=0}^{\omega(i)}
		&
		\left [
		\left(
		\begin{matrix}
			E_{i'j',ij} & 0\\
			0 & E_{i'j',ij}
		\end{matrix}
		\right) \right.\nonumber\\
		+&
		\left. \frac{d_{\rm eff}}{2\pi \lambda^2}\left(
		\begin{matrix}
			bK_{i'j',ij}^{11} & aK_{i'j',ij}^{12}\\
			bK_{i'j',ij}^{21} & aK_{i'j',ij}^{22}
		\end{matrix}
		\right) \right ] 
		\left(
		\begin{matrix}
			u_{ij\alpha}^1\\u_{ij\alpha}^2
		\end{matrix}
		\right)=
		\left(
		\begin{matrix}
			F_{i'j'\alpha}^1\\F_{i'j'\alpha}^2
		\end{matrix}
		\right)
		\label{eq:5}.
	\end{align}
	The elements of coefficient matrix can be calculated as follows:
	\begin{eqnarray}
		&E_{i'j',ij}=	\left \langle \ell_{i'j'},\omega_{ij}(x,y) \right \rangle,\nonumber\\
		&K_{i'j',ij}^{11}=\left \langle \ell_{i'j'},(K\omega_{ij})\left(x,y,\frac{b}{a}\right) \right \rangle,\nonumber\\
		&K_{i'j',ij}^{12}=\left \langle \ell_{i'j'},(K\omega_{ij})\left(y,x,\frac{a}{b}\right) \right \rangle,\nonumber\\
		&K_{i'j',ij}^{21}=\left \langle \ell_{i'j'},(K\omega_{ij})\left(y,x,\frac{b}{a}\right) \right \rangle,\nonumber\\
		&K_{i'j',ij}^{22}=\left \langle \ell_{i'j'},(K\omega_{ij})\left(x,y,\frac{a}{b}\right) \right \rangle,\nonumber\\
		& F_{i'j'\alpha}^1=	\left \langle \ell_{i'j'},f_\alpha(x,y) \right \rangle,\nonumber\\
		& F_{i'j'\alpha}^2=	\left \langle \ell_{i'j'},f_\alpha(y,x) \right \rangle
		\label{eq:6},
	\end{eqnarray}
	where $\left \langle \ell,F \right \rangle$ represents the value of functional $\ell$ evaluated at function $F$ (see Appendix \ref{sec:appendix A2}). 
	\par 
	Owing to the tight support properties of $\omega_{ij}$,  $\{E_{i'j',ij}\}$is an upper-triangular sparse matrix. Whereas for $\left\{K_{i'j',ij}^{\beta\gamma}\right\}\,(\beta=1,2,\,\gamma=1,2)$, the number of non-zero elements or equivalent two-dimensional singular integers to be calculated is $\sim 10^8$, which would cost considerable computation time. Fortunately, it was established that $\left\{K_{i'j',ij}^{\beta\gamma}\right\}$ can be approximated by a compressed sparse matrix, $\left\{\widetilde{K_{i'j',ij}^{\beta\gamma}}\right\}$\cite{Xu-2005}. A detailed compression algorithm is presented in Appendix \ref{sec:appendix B}. After constructing the coefficient matrix, $u_{ij\alpha}^\beta$ can be determined by solving Eq.~(\ref{eq:5}). Subsequently, the screening current density can be obtained according to Eq.~(\ref{eq:4}).
	\par 
	Finally, the mutual inductance, $M_{\rm cal}$, is calculated by integrating the vector potential of the drive and screening currents around each loop of the pickup coil. In this study, we extracted the absolute $\lambda$ by interpolating $M_{\rm exp}/M_{\rm exp} (T>T_{\rm c})$ into a “lookup” table consisting of $M_{\rm cal}/M_{\rm cal}(T>T_{\rm c})$ for different $\lambda$ values.

	\section{Results and discussion}
\begin{figure}[ht!]
		\centering
		\includegraphics[width=0.4\textwidth]{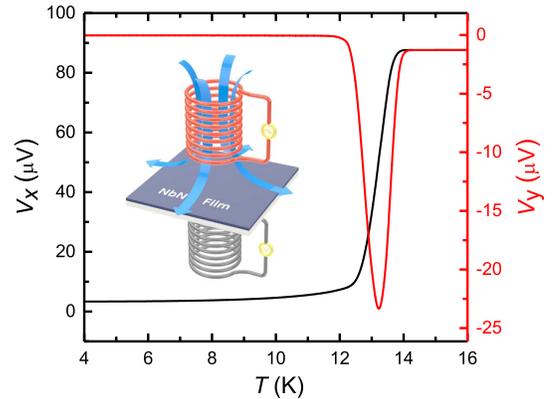}
		\caption{\label{fig1} Temperature dependence of the pickup coil voltage for NbN\#1. The black and red curves represent the real and imaginary components of the pickup coil voltage, respectively. The inset shows the schematic illustration of the MI device.}
	\end{figure}

	We measured NbN\#1 and NbN\#2 using our home-made \emph{transmission}-type MI device. The raw data for NbN\#1 is shown in Fig.~\ref{fig1}. 
	It is established that  strong diamagnetic screening emerges when the sample enters the Meissner state at $T_{\rm c}\sim 14 \, \rm K$, which is reflected as a sudden drop of $V_x$ in the pickup coil (black line). Correspondingly, $V_y$ shows a clear dip (red line), which may be attributed to energy dissipation mechanisms such as vortex-antivortex unbinding \cite{Hebard-1980,Leemann-1986}.
	\subsection{Breakdown of FLS method}
\begin{figure}[ht!]
		\centering
		\includegraphics[width=0.38\textwidth]{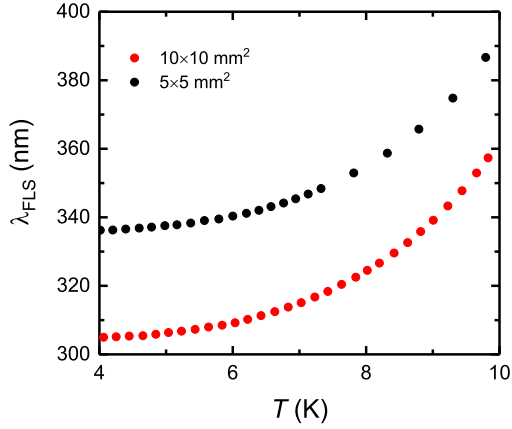}
		\caption{\label{fig2} Temperature-dependent penetration depth, $\lambda_{\rm FLS}$, of NbN\#1 (red circles) and NbN\#2 (black circles) extracted using the FLS method, showing non-negligible deviation at low temperature.}
	\end{figure}

	We first employed the FLS method to extract $\lambda$ (denoted as $\lambda_{\rm FLS}$) of NbN\#1 ($10\times 10 \, \rm mm^2$) and NbN\#2 ($5\times 5 \, \rm mm^2$) from the mutual inductance, the details are described in \cite{Zhang-2020}.  Notably, the $\lambda_{\rm FLS}$ values of the two films exhibit significant discrepancies at low temperatures (Fig.~\ref{fig2}). The $\lambda_{\rm FLS} \, (T=4 \, \rm K)$ of NbN\#2 is $\sim10\%$ higher than that of NbN\#1. This phenomenon prompts us to recall the criterion given by Turneaure \emph{et al.} \cite{Turneaure-1996}, which predicts that the FLS method fails when the film size is less than about five times the diameter of the coils. That is because the screening current reaches a significant peak at the film edge for small-sized films, which may cause a non-negligiable contribution to $M_1$. Thus $M_1$ depends on $\lambda$, leading to the failure of the FLS method. Considering that the outer diameters of our coils are $1.3 \, \rm mm$, we speculate that the FLS method may be invalid for the NbN\#2 with a size of $5\times 5 \, \rm mm^2$. 
	\par 
	\subsection{Calculated screening currents}
\begin{figure}[ht!]
		\centering
		\includegraphics[width=0.47\textwidth]{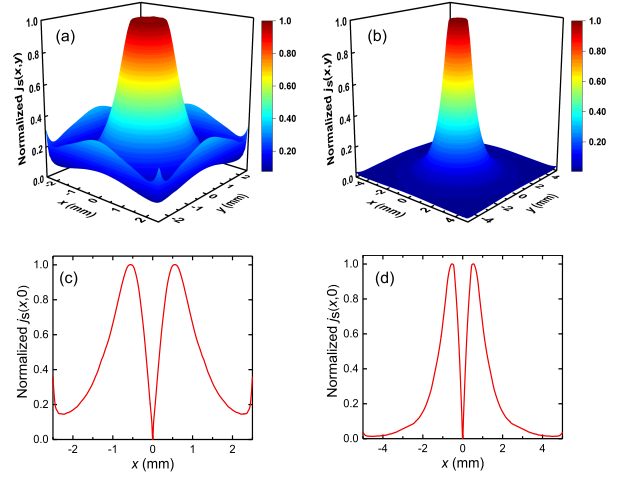}
		\caption{\label{fig3}Normalized screening current densities $j_s=\sqrt{j_{sx}^2+j_{sy}^2}$ for $5\times 5 \, \rm mm^2$ and $10\times 10 \, \rm mm^2$  superconducting films, calculated by the FWC method. (a-b) Three-dimensional false-color plot of the normalized $j_s(x,y)$. (c-d) Cuts of (a) and (b) at $y=0$. }
	\end{figure}

	To verify our speculation, we calculated the screening currents for square superconducting films with $d=6.5 \, \rm nm$ and $\lambda=300\, \rm nm$ by solving Eq.~(\ref{eq:2}). Fig.~\ref{fig3} shows the normalized screening current densities, $j_s=\sqrt{j_{sx}^2+j_{sy}^2}$, for the $5\times 5 \, \rm mm^2$ and $10\times 10 \, \rm mm^2$ films. They both attain a local maximum around the radius of the drive coil, which is consistent with previous reports \cite{Turneaure-1996,Turneaure-1998}. Notably, there indeed exists a significant peak at the edge of the $5\times 5 \, \rm mm^2$ film. Thus, we conclude that the breakdown of the FLS method for small-sized films is due to the large screening current at the film edge, which is a crucial finding in this study.
	\par 
	\subsection{$\lambda$ re-extracted using the FWC method}
	In contrast, the screening current  at the film edge is considered in our numerical model, so the FWC method in principle works for small-sized films. 
	\begin{figure}[ht!]
		\centering
		\includegraphics[width=0.47\textwidth]{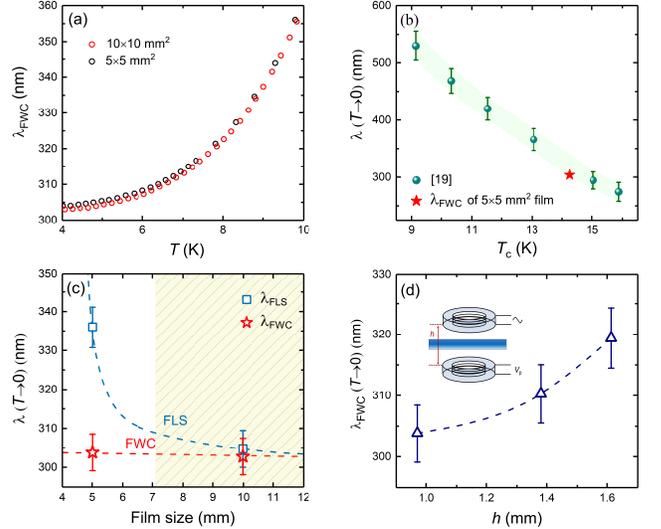}
		\caption{\label{fig4}(a) Temperature-dependent $\lambda_{\rm FWC}$ of NbN\#1 (red circles) and NbN\#2 (black circles) extracted using the FWC method. (b) Value of $\lambda_{\rm FWC} (T\to 0)$ for NbN\#2 extracted using the FWC method, which shows a good agreement with the published value \cite{Kamlapure-2010}; the length of error bar is shorter than the symbol size. (c) Film-size-dependent penetration depth extracted using the FLS and FWC methods. The shadow region indicates where the FLS method works \cite{Turneaure-1996}; the dashed line is a visual guide.  (d) Measurements of $\lambda_{\rm FWC} (T\to 0)$ for NbN\#2 with different coil spacings. }
	\end{figure}
	To elucidate this, we re-extracted $\lambda$ (denoted as $\lambda_{\rm FWC}$) based on the “lookup” tables depicted in Section \ref{sec:level3}. As shown in Fig.~\ref{fig4}(a), the low-temperature data for the NbN\#1 and NbN\#2 are consistent.
	The difference in film size resulted in a deviation of only $\sim 1\, \rm nm$ in $\lambda_{\rm FWC} \, (T=4 \, \rm K)$. Moreover, the values of the extrapolated zero-temperature penetration depth $\lambda_{\rm FWC}(0)$ are consistent with the data in literature \cite{Kamlapure-2010} (see Fig.~\ref{fig4}(b)).
	\par 
	To summarize, we present the film-size-dependent $\lambda$ extracted using the FLS and FWC methods in Fig.~\ref{fig4}(c). It is established that for large-sized films (see the shadow region where the “Turneaure criterion” meets),  $\lambda_{\rm FLS}$ is almost similar to $\lambda_{\rm FWC}$. However, as the film size decreases, $\lambda_{\rm FLS}$ changes significantly whereas $\lambda_{\rm FWC}$ changes slightly. This indicates the applicability of the FWC method for small-sized superconducting films.
	\par 
	In addition, we tested the tolerance of the FWC method to different coil spacings. As shown in Fig.~\ref{fig4}(d),
	the variation of $\lambda_{\rm FWC}$ is only 5\% when the coil spacing reaches 1.6 mm. This spacing is sufficiently large for the ionic liquid gating experiment, which is a powerful tool for manipulating the superconducting properties continuously  \cite{Goldman-2014,Qin-2020}.
	\section{Conclusions}
	In summary, we propose the FWC method for the MI technique to extract the absolute $\lambda$ of polygonal superconducting films with arbitrary sizes. The experimental results on the square NbN films indicate that the absolute $\lambda$ extracted using the FWC method is independent of the film size, whereas the conventional FLS method fails for the $5\times 5 \, \rm mm^2$ film because of the large screening current at the film edge. This numerical method allows us to directly determine $\lambda$ of small-sized superconducting films, dispensing with extra manipulations. In addition, for a coil spacing of 1.6 mm, the error in $\lambda_{\rm FWC}$ is only $\sim 5\%$. The high tolerance to coil spacing is promising in the \emph{in situ} $\lambda$ measurements integrated with ionic liquid gating technique, paving a high-efficiency way for the superconductivity research.
	
	\begin{acknowledgments}
		This work was supported by the Strategic Priority Research Program (B) of Chinese Academy of Sciences (XDB25000000), Key-Area Research and Development Program of Guangdong Province (2020B0101340002), the National Key Basic Research Program of China (2017YFA0302902, 2017YFA0303003 and 2018YFB0704102), the National Natural Science Foundation of China (11927808, 11834016, 118115301, 119611410 and 11961141008), the Key Research Program of Frontier Sciences, CAS (QYZDB-SSW-SLH008 and QYZDY-SSW-SLH001), CAS Interdisciplinary Innovation Team, and Beijing Natural Science Foundation (Z190008).
	\end{acknowledgments}
	\appendix
	\section{The multi-scale bases and collocation functionals}
	
	\subsection{\label{sec:appendix A1} Multi-scale wavelets}
	We select the two-dimensional linear multi-scale wavelets $\omega_{ij}(x,y) \,((x,y)\in E=\{(x,y)\in \mathbb R^2:0\leq x\leq y\leq 1\})$ in \cite{Chen-2008} to expand the screening currents, where subscript $i$ represents the level of $\omega_{ij}$ and $j$ represents the serial number of $\omega_{ij}$ at the $i$th level. For convenience, let $\mathbb Z_n=\{0,1,\cdots,n-1\}$, and $\mathbb Z_n^m=\mathbb Z_n\times\mathbb Z_n\times\cdots\times \mathbb Z_n$. We define a family of contract mappings $\Phi=\{\phi_i:i\in\mathbb Z_4\}$ with $\phi_0=(x/2,y/2), \phi_1=(x/2,(y+1)/2), \phi_2=((1-x)/2,1-y/2), \phi_3=((x+1)/2,(y+1)/2)$.
	\par 
	The three wavelets at level 0 are constructed as 
	\begin{eqnarray}
		\omega_{00}(x,y)=-3x+2y,\nonumber\\
		\omega_{01}(x,y)=2+x-3y,\nonumber\\
		\omega_{02}(x,y)=-1+2x+y
		\label{eq:A1}.
	\end{eqnarray}
	The nine wavelets at level 1 are given by
	\begin{eqnarray}
		\omega_{10}(x,y) &=&
		\begin{cases}
			-\frac{11}{8}-\frac{15}{8}x+\frac{41}{8}y       & (x,y)\in S_0, \\
			\frac{5}{8}+\frac{1}{8}x-\frac{7}{8}y& (x,y)\in E\setminus S_0,
		\end{cases}\nonumber\\
		\omega_{11}(x,y) &=&
		\begin{cases}
			1-\frac{15}{4}x-\frac{7}{8}y       & (x,y)\in S_0, \\
			-1+\frac{1}{4}x+\frac{9}{8}y& (x,y)\in E\setminus S_0,
		\end{cases}\nonumber\\
		\omega_{12}(x,y) &=&
		\begin{cases}
			\frac{9}{8}+\frac{15}{8}x-\frac{29}{8}y       & (x,y)\in S_0, \\
			-\frac{15}{8}-\frac{1}{8}x+\frac{19}{8}y& (x,y)\in E\setminus S_0,
		\end{cases}\nonumber\\
		\omega_{13}(x,y) &=&
		\begin{cases}
			-\frac{15}{8}-\frac{41}{8}x+\frac{13}{4}y       & (x,y)\in S_1, \\
			\frac{1}{8}+\frac{7}{8}x-\frac{3}{4}y& (x,y)\in E\setminus S_1,
		\end{cases}\nonumber\\
		\omega_{14}(x,y) &=&
		\begin{cases}
			\frac{29}{8}+\frac{7}{8}x-\frac{37}{8}y       & (x,y)\in S_1, \\
			-\frac{3}{8}-\frac{9}{8}x+\frac{11}{8}y& (x,y)\in E\setminus S_1,
		\end{cases}\nonumber\\
		\omega_{15}(x,y) &=&
		\begin{cases}
			-\frac{5}{8}-\frac{29}{8}x+\frac{7}{4}y       & (x,y)\in S_1, \\
			\frac{3}{8}+\frac{19}{8}x-\frac{9}{4}y& (x,y)\in E\setminus S_1,
		\end{cases}\nonumber\\
		\omega_{16}(x,y) &=&
		\begin{cases}
			\frac{15}{4}-\frac{13}{4}x-\frac{15}{8}y       & (x,y)\in S_3, \\
			-\frac{1}{4}+\frac{3}{4}x+\frac{1}{8}y& (x,y)\in E\setminus S_3,
		\end{cases}\nonumber\\
		\omega_{17}(x,y) &=&
		\begin{cases}
			-\frac{1}{8}-\frac{37}{8}x+\frac{15}{4}y       & (x,y)\in S_3, \\
			-\frac{1}{8}+\frac{11}{8}x-\frac{1}{4}y& (x,y)\in E\setminus S_3,
		\end{cases}\nonumber\\
		\omega_{18}(x,y) &=&
		\begin{cases}
			-\frac{5}{2}+\frac{7}{4}x+\frac{15}{8}y       & (x,y)\in S_3, \\
			\frac{1}{2}-\frac{9}{4}x-\frac{1}{8}y& (x,y)\in E\setminus S_3,
		\end{cases}
	\end{eqnarray}
	where $S_i  =\phi_i (E)$ $(i \in \mathbb Z_4)$ and $E\setminus S_i$ represents the set difference of $S_i$ from $E$.
	\par 
	To generate high level wavelets $\omega_{ij}$ for $i\geq2$, we introduce the composite map $\phi_{{\bf e}}=\phi_{e_0}\circ \cdots \circ \phi_{e_{n-1}}$, where ${\bf e}=(e_0,\cdots.e_{n-1})\in \mathbb Z_4^n$ is an $n$-dimensional vector, and a number associated with ${\bf e}$ is defined as $\mu({\bf e})=\sum_{k=1}^{n}4^{n-k}e_{k-1}$. Additionally, we define operators $\mathcal{T}_e \,(e\in \mathbb Z_4)$ by $\mathcal{T}_eF(x,y)=F(\phi_e^{-1}(x,y))\chi_{S_e}(x,y)$, where $\chi_{S_e}$ denotes the characteristic function of set $S_e$. The corresponding composite operator is $\mathcal{T}_{\bf e}=\mathcal{T}_{e_0}\circ \cdots \mathcal{T}_{e_{n-1}}$ . Thereafter, the high-level multiscale wavelets, $\omega_{ij}\,(i\geq 2)$,  are constructed as
	\begin{equation}
		\omega_{ij}=\mathcal{T}_{{\bf e}}\omega_{1l},\quad j=9\mu({\bf e})+l,\,{\bf e}\in\mathbb Z_4^{i-1},l\in\mathbb Z_9.
	\end{equation}
	\subsection{\label{sec:appendix A2}Collocation functionals}
	Using collocational functionals $\ell_{ij}$, we can discretize Eq.~(\ref{eq:3}) into Eq.~(\ref{eq:5}), which is easier to handle. The concrete form of $\ell_{ij}$ is given in \cite{Chen-2008}. The three functionals of level 0 are given by
	\begin{equation}
		\ell_{0j}=\delta_{t_{0j}},\quad j\in \mathbb Z_4,
	\end{equation}
	where $\delta_{t_{0j}}$ is the point evaluation functionals of three points $t_{00}=(1/7,5/7)$, $t_{01}=(2/7,3/7)$, $t_{02}=(4/7,6/7)$, and its effect on function $F$ is defined as $\left \langle \delta_{t_{0j}},F\right \rangle=F(t_{0j})$. The nine collocation functionals at level 1 are constructed as 
	\begin{equation}
		\ell_{1l}=\sum_{e\in \mathbb Z_{12}}^{}c_{le}\delta_{t_{1e}},\quad l\in \mathbb Z_9,
	\end{equation}
	where $t_{1e}=\phi_i (t_{0j})\, (e=3i+j,i\in \mathbb Z_4,j\in \mathbb Z_3)$ and the matrix $C=\{c_{le}\}_{9\times12}$ is constructed as
	\begin{equation}
		C=
		\left(                 
		\begin{array}{cccccccccccc}   
			0& 1& 0& 0 &1& 0& 0 &-1 &0& 0& 0&-1\\
			0 &-1 &0 &0& 1& 1& 0& 0& 0 &0& 0&-1\\
			0& 1& 0& 0& 0& -1& 0& 0& 1& 0& 0 &-1\\
			-1 &0& 0& 1& 0& 0& 0& 0 &1 &-1& 0& 0\\
			1& 0& 0 &0& 0& 0& 0 &0& 1 &-1& 0 &-1\\
			1 &0 &0& 0& 0& 1& 0& 0& -1& 0& 0& -1\\
			0 &0 &1 &0& 0& -1& 1& 0& 0& 0& -1& 0\\
			0& 0& 1& 0& 0& 1 &-1& 0& -1& 0& 0& 0\\
			0& 0& 1& 0& 0 &-1& 0& 0& -1 &0 &0& 1\\
		\end{array}
		\right) 
		.
	\end{equation}
	\par 
	Subsequently, we introduce a linear operator, $\mathcal{L}_e$, defined for functional $\ell$ and function $F$ by the equation $\left \langle \mathcal{L}_e \ell,F\right \rangle=\left \langle \ell,F\circ \phi_e\right \rangle$, and its composite operator $\mathcal{L}_{{\bf e}}=\mathcal{L}_{e_0}\circ \cdots \circ \mathcal{L}_{e_{n-1}}$. With the initial functionals, we can construct collocation functionals for $i\geq 2$ as
	\begin{equation}
		\ell_{ij}=\mathcal{L}_{{\bf e}}\ell_{1l},\quad j=9\mu({\bf e})+l,\,{\bf e}\in\mathbb Z_4^{i-1},\,l\in\mathbb Z_9.
	\end{equation}
	\section{\label{sec:appendix B}Block truncation scheme}
	Using block truncation schemes presented in \cite{Xu-2005}, we can approximate matrix $\left\{K_{i'j',ij}^{\beta\gamma}\right\}\,(\beta=1,2,\,\gamma=1,2)$ by a sparse matrix $\left\{\widetilde{K_{i'j',ij}^{\beta\gamma}}\right\}$. According to Eq.~(\ref{eq:6}), the computation of matrix element $K_{i'j',ij}^{\beta\gamma}$ is associated with the corresponding $\ell_{i'j'}$ and $\omega_{ij}$, which are further associated with a unique pair of vectors ${\bf e}'$ and ${\bf e}$ such that $\ell_{i'j'}=\mathcal{L}_{{\bf e}'}\ell_{1l'}$ and $\omega_{ij}=\mathcal{T}_{{\bf e}}\omega_{1l}$. Hence, $K_{i'j',ij}^{\beta\gamma}$ corresponds to a unique pair of ${\bf e}'$ and ${\bf e}$.
	\par 
	In addition, the contractive mapping $\phi_{{\bf e}}$ can map the unit triangle $E$ into a smaller triangle $S_{ij} (E)$, which has a centroid $(C_x,C_y)$. Therefore, we define $\Gamma({\bf e})=4^{(i-1)/2}(C_x,C_y)$ and assign another pair of vectors ${\bf q}'$ and ${\bf q}$ to the matrix element $K_{i'j',ij}^{\beta\gamma}$ as follows:
	\begin{enumerate}
		\item[1)] 	If $i\geq i'$, set ${\bf e}_c=(e_0,e_1,\cdots,e_{i'-2})$ and ${\bf q}'=\Gamma({\bf e}')$, ${\bf q}=\Gamma({\bf e}_c)$.
		\item[2)] 	If $i< i'$, set ${\bf e}_c'=(e_0',e_1',\cdots,e_{i'-2}')$ and ${\bf q}'=\Gamma({\bf e}_c')$, ${\bf q}=\Gamma({\bf e})$.
	\end{enumerate}
	\par 
	Consequently, $\{K_{i'j',ij}^{\beta\gamma}\}$ can be divided into blocks $\{K_{{\bf q}',{\bf q}}^{\beta\gamma}\}$ according to ${\bf q}'$ and ${\bf q}$ as
	$
	K_{{\bf q}',{\bf q}}^{\beta\gamma}=\{K_{i'j',ij}^{\beta\gamma}: K_{i'j',ij}^{\beta\gamma} \text{corresponds to the same pair of ${\bf q}'$ and ${\bf q}$}  \}
	$. Given the truncation parameter $\sqrt{2}\leq r_{i'i}\leq \sqrt{3}$, the block truncation scheme indicates that $K_{{\bf q}',{\bf q}}^{\beta\gamma}$ can be replaced by
	\begin{eqnarray}
		\widetilde{K_{{\bf q}',{\bf q}}^{\beta\gamma}} &=&
		\begin{cases}
			K_{{\bf q}',{\bf q}}^{\beta\gamma}       & |{\bf q}-{\bf q}'|\leq r_{i'i}, \\
			0& \text{otherwise}.
		\end{cases}
	\end{eqnarray}

\bibliography{ref} 

\end{document}